\documentclass[aps,prd,reprint,superscriptaddress,nofootinbib,floatfix]{revtex4-2}
\usepackage{amsmath,amssymb,bm,graphicx,booktabs,array,placeins,slashed}
\usepackage[T1]{fontenc}
\usepackage{microtype}
\usepackage[colorlinks=true,linkcolor=blue,citecolor=blue,urlcolor=blue]{hyperref}

\makeatletter
\renewcommand{\paragraph}{%
  \@startsection{paragraph}{4}{\parindent}{.5\baselineskip}{-1em}%
    {\normalfont\normalsize\itshape}%
}
\makeatother

\newcommand{\bibtolerant}{\sloppy\hbadness=10000\relax}

\newcommand{\dd}{\mathrm d}
\newcommand{\kb}{\boldsymbol k_\perp}
\newcommand{\Db}{\boldsymbol\Delta_\perp}
\newcommand{\qb}{\boldsymbol q_\perp}
\newcommand{\FS}{\mathcal F_S}
\newcommand{\FA}{\mathcal F_A}
\newcommand{\FK}{\mathcal F_{Dk}}

\newcommand{\FD}{\mathcal F_{D\Delta}}
\newcommand{\GS}{\mathcal G_S}
\newcommand{\GA}{\mathcal G_A}
\newcommand{\GK}{\mathcal G_{Dk}}

\newcommand{\GD}{\mathcal G_{D\Delta}}
\newcommand{\Ipol}{\mathcal I}
\newcommand{\Vpol}{\mathcal V}
\newcommand{\Tpol}{\mathcal T}
\newcommand{\TTpol}{\widetilde{\mathcal T}}
\newcommand{\Cf}{\mathcal C_f}
\newcommand{\af}{a_f}
\newcommand{\id}{\boldsymbol 1}
\newcommand{\Lcut}{\mathcal L}
\newcommand{\Rcut}{\mathcal R}
\newcommand{\Luv}{\Lambda_{\rm UV}}

\newcommand{\PaperTitle}{Generalized transverse-momentum-dependent parton
distributions of a photon and orbital angular momentum}

\providecommand{\Leqref}[1]{\eqref{#1}}
 \hypersetup{
  pdftitle={\PaperTitle},
  pdfauthor={Sujit Jana, Asmita Mukherjee, Sreeraj Nair},
  pdfsubject={Physical Review D Letter with Supplemental Material},
  hypertexnames=false
}

\begin{document}
\title{Generalized transverse-momentum-dependent parton distributions  
\texorpdfstring{\\}{ }of a photon and orbital angular momentum}
\author{Sujit Jana}
\affiliation{Department of Physics, Indian Institute of Technology Bombay, Powai, Mumbai 400076, India}
\author{Asmita Mukherjee}
\affiliation{Department of Physics, Indian Institute of Technology Bombay, Powai, Mumbai 400076, India}
\author{Sreeraj Nair}
\affiliation{Frontiers Science Center for Rare Isotopes, and School of Nuclear Science and Technology,
Lanzhou University, Lanzhou 730000, China}
\date[]{}

\begin{abstract}
The coefficients of generalized transverse-momentum-dependent distributions
(GTMDs) that encode the canonical fermion orbital angular momentum and
spin--orbit correlation do not contribute to the direct
transverse-momentum-dependent (TMD) and generalized-parton-distribution
(GPD) projections. We present, to our knowledge, the first explicit
calculation of the leading-twist, time-even, chiral-even fermion GTMDs of an
on-shell real photon at zero skewness and order~$\alpha$. The coefficients
$\mathcal{F}_A$ and $\mathcal{G}_A$, which multiply the
$\bm{k}_\perp\times\bm{\Delta}_\perp$ structure, yield the fermion orbital
angular momentum $\ell_z^f$ and spin--orbit correlation $\mathcal{C}_z^f$.
For photon helicity $+1$, both are obtained in closed form in the leading
fermion--antifermion sector and obey $\mathcal{C}_z^f=-\ell_z^f/2$ for a
common transverse regulator independent of $x$, because the channel with the
active fermion's spin anti-aligned with its orbital motion carries three
times the integrated weight of the aligned channel. The projections reproduce the
known photon TMDs and leading-logarithmic photon-to-fermion kernels, and
energy-momentum-tensor overlaps reproduce the known one-loop photon
gravitational form factors $F_1$, $F_2$, and $F_3$. Our result provides an
analytic benchmark for fermionic orbital and spin--orbit structure in the
real photon.
\end{abstract}
\maketitle
\setlength{\parskip}{0pt plus 0pt minus 0pt}
\flushbottom

\paragraph*{Introduction.}

The real photon's fermion--antifermion component is perturbatively
calculable \cite{witten1977anomalous}. Its structure
functions and polarized parton distribution functions (PDFs)
\cite{Nisius2000photon,krawczyk2001photon,StratmannVogelsang1996photon,
bass1998photonspin}, generalized parton distributions (GPDs)
\cite{friot2007photongpd,mukherjee2011generalized,mukherjee2012nonzerozeta,
mukherjee2013generalized,gabdrakhmanov2012photongpd}, and
transverse-momentum-dependent distributions (TMDs)
\cite{Nair:2022evk,Nair:2023lir,puhan2025photontmd,puhan2025photonlfwf}
resolve its partonic structure.
Gravitational form factors (GFFs) describe its momentum,
angular-momentum, and stress distributions \cite{freese2022spatial},
with related studies for a dressed quark
\cite{more2022gravitational,more2023sumrules}.

Generalized transverse-momentum-dependent distributions (GTMDs) depend
jointly on the parton transverse momentum $\kb$ and momentum transfer $\Db$
\cite{meissner2009gtmd,lorce2013gtmd}. Setting $\Db=0$ gives TMDs,
integrating over $\kb$ gives GPDs, and Fourier transformation in $\Db$
gives Wigner distributions
\cite{ji2003wigner,belitsky2004wigner,lorce2011unified}.
Parton orbital angular momentum (OAM) depends on its operator definition
\cite{jaffe1990spin,ji1997gpd,leader2014angular}.
For spin-$\tfrac12$ targets, $k_\perp^2$ moments of $F_{1,4}$ and $G_{1,1}$,
respectively, give OAM and spin--orbit correlation
\cite{lorce2011overlap,lorce2012oam,hatta2012notes,
ji2012oam,lorce2014spinorbit,kanazawa2014gtmd}.
Kinetic OAM and spin--orbit correlation also follow from GPD moments
\cite{ji1997gpd,lorce2014spinorbit}.

Proposed exclusive measurements probe nucleon GTMDs
\cite{hatta2016dijet,bhattacharya2017ddy,bhattacharya2022gluonoam},
and lattice calculations compute the relevant moments
\cite{engelhardt2017oam}. Spin--orbit correlations
have been extracted from lattice-QCD GPD moments
\cite{bhattacharya2025lattice}, and their measurement has been proposed
for the Electron-Ion Collider \cite{bhattacharya2025eic}.
They have also been analyzed in the twist expansion
\cite{hatta2024twist}, and computed for spin-0 and spin-1 hadrons and
light-front two-body systems
\cite{lorce2025spin0,choudhary2026pion,acharyya2024spin1,hu2026quarkonium}.
Light-front wave-function overlaps relate these moments to the orbital
and helicity content of individual amplitudes
\cite{lorce2011overlap,mukherjee2014wigner,more2017wigner,
more2018gluonwigner}.

A spin-1 target admits tensor-polarized structures
\cite{bacchetta2000spinone,kumano2021spinone,cosyn2019spinone}, including a
tensor-polarized spin--orbit correlation \cite{kim2026spin1}.
The real photon's two transverse polarizations allow helicity conservation
and helicity flip by two units.

In this Letter we derive the leading-twist, time-even, chiral-even fermion
GTMDs of a real photon at zero skewness and order $\alpha$ from its
leading fermion--antifermion light-front wave function. We identify the
GTMDs that carry the canonical OAM and spin--orbit correlation
and resolve their contributions by fermion helicity. We then obtain the
TMD, GPD, and PDF projections. GTMD sum rules determine two photon GFFs,
while the transverse energy-momentum tensor gives the third.

\paragraph*{Kinematics and light-front wave function.}
\setlength{\parskip}{0pt plus 0pt minus 0pt}

We use the symmetric zero-skewness frame
\begin{equation}
 \begin{gathered}
 P=\frac{p'+p}{2},\quad \Delta=p'-p,\quad t=\Delta^2=-\Delta_\perp^2,\\
 p^2=p'^2=M_\gamma^2=0,\quad \Delta^+=0,\quad \boldsymbol P_\perp=0.
 \end{gathered}
 \label{eq:kinematics}
\end{equation}
Here $p$ and $p'$ are the initial and final photon momenta. Our light-front
conventions are $v^\pm=v^0\pm v^3$ and
$v\cdot w=(v^+w^-+v^-w^+)/2-\boldsymbol v_\perp\cdot\boldsymbol w_\perp$.
We retain the following two-sector Fock expansion of the photon:
\begin{equation}
 \begin{aligned}
 |\gamma,h\rangle
 =\sqrt{Z_\gamma}\bigg[
 &|\gamma,h\rangle_1
 +\sum_{s,\bar s}\int\dd x\,\dd^2q_\perp\,
 \Psi^h_{s\bar s}(x,\qb)\\[-2pt]
 &\quad\times|f\bar f;x,\qb,s,\bar s\rangle_2\bigg].
 \end{aligned}
 \label{eq:state}
\end{equation}
The first sector is a bare photon and the second a fermion--antifermion
pair; $Z_\gamma$ fixes the state normalization. Here $h=\pm1$ is the photon
helicity, and $s/2$ and $\bar s/2$ are the fermion and antifermion
light-front helicities, with $s,\bar s=\pm1$. The variables
$x=k^+/P^+\in(0,1)$ and $\qb$ are the fermion momentum fraction and relative
transverse momentum. The wave
function~\cite{harindranath1999radiative,mukherjee2011generalized,mukherjee2013generalized,
lappi2020relativistic,beuf2016dipole} is
\begin{equation}
 \begin{gathered}
 \Psi^h_{s\bar s}(x,\qb)
 =-\frac{e e_f}{\sqrt{2(2\pi)^3}}\,
 \frac{\chi_s^\dagger \mathcal K_h(\qb)\chi_{-\bar s}}
 {\qb^2+m^2},\\
 \mathcal K_h(\qb)
 =\left[(\beta\,\boldsymbol\sigma\cdot\qb-im)\sigma^i
 -x\sigma^i(\boldsymbol\sigma\cdot\qb)\right]\epsilon_\perp^{h\,i}.
 \end{gathered}
 \label{eq:wavefunction}
\end{equation}
Here $m>0$ and $ee_f$ are the fermion mass and charge, $\beta=1-x$,
$\boldsymbol\epsilon_\perp^h=(-h,-i)/\sqrt2$, and
$\chi_+=(1,0)^T$, $\chi_-=(0,1)^T$ are two-component spinors.
The $\sigma^i$ are Pauli matrices, with the transverse index $i=1,2$ summed.
Each helicity amplitude has relative orbital angular-momentum projection
$\ell_z=h-(s+\bar s)/2$ \cite{brodsky2001lcspin}.

\paragraph*{Fermion GTMDs of the photon.}

The correlator and its leading $f\bar f$ overlap are
\begin{equation}
 \begin{aligned}
 W_{h'h}^{[\Gamma]}
 &=\frac14\int\frac{\dd z^-\,\dd^2z_\perp}{(2\pi)^3}
 e^{ik\cdot z}\,
 \langle\gamma(p'),h'|\\
 &\quad\times\bar\psi(-z/2)\mathcal U\Gamma\psi(z/2)
 |\gamma(p),h\rangle\big|_{z^+=0}\\
 &=N_c\sum_{s',s,\bar s}
 \Psi^{h' *}_{s'\bar s}(x,\boldsymbol q_+)
 S^{[\Gamma]}_{s's}\,
 \Psi^h_{s\bar s}(x,\boldsymbol q_-).
 \end{aligned}
 \label{eq:overlap}
\end{equation}
Here $\psi$ is the fermion field, $z$ the field separation, and
$\Gamma=\gamma^+$ or $\gamma^+\gamma_5$ the Dirac projection.
The Wilson line $\mathcal U$ is taken as unity in light-cone gauge
$A^+=0$ with $A_\perp(+\infty)=0$.
The factor $N_c$ equals $1$ for a lepton and $3$ for a quark flavor.
For light quarks, this describes the pointlike component of the photon;
its nonperturbative hadronic component is not included.
The primed helicities label the final state; the spectator helicity
$\bar s/2$ is unchanged.
The relative transverse momenta are $\boldsymbol q_\pm=\kb\pm\beta\Db/2$
in the final ($+$) and initial ($-$) wave functions.
We evaluate the bare two-body overlap through $O(\alpha)$, with
$\alpha=e^2/(4\pi)$.
The two-component reduction of the
currents \cite{zhang1993twocomponent} gives the active-fermion spin matrices
$S^{[\gamma^+]}=\id$ and $S^{[\gamma^+\gamma_5]}=\sigma^3$.
The Supplemental Material (SM), appended after the references, gives the
helicity amplitudes and overlap details.

The trace, antisymmetric part, and symmetric traceless part of the photon's
$2\times2$ polarization matrix give scalar, vector, and tensor sectors
\cite{bacchetta2000spinone}.
We use the polarization structures
\begin{equation}
 \begin{gathered}
 \Ipol_{h'h}=B^{ii}_{h'h},\qquad
 \Vpol_{h'h}=-i\varepsilon_\perp^{ij}B^{ij}_{h'h},\\
 \Tpol^{ij}_{h'h}=2B^{(ij)}_{h'h}-\delta^{ij}\Ipol_{h'h},\quad
 \TTpol^{ij}_{h'h}
 =i\varepsilon_\perp^{ik}\Tpol^{kj}_{h'h}.
 \end{gathered}
 \label{eq:tensors}
\end{equation}
Here $B^{ij}_{h'h}=\epsilon_\perp^{h'\,*i}\epsilon_\perp^{h\,j}$, with
final helicity $h'$ and initial helicity $h$. Parentheses denote
symmetrization, $2B^{(ij)}_{h'h}=B^{ij}_{h'h}+B^{ji}_{h'h}$;
transverse indices run over $1,2$ and repeated indices are summed,
$\delta^{ij}$ is the unit tensor, and $\varepsilon_\perp^{12}=1$.
We write $\kb\times\Db=k_x\Delta_y-k_y\Delta_x$.
The structures $\Ipol$, $\Vpol$, and $\Tpol$ describe scalar, circular, and
linear polarization, respectively. The dual $\TTpol$ supplies the axial
tensor structure with the same polarization components. The SM gives the
decomposition and tensor identities. For $h'=h$, $\Ipol_{hh}=1$ and
$\Vpol_{hh}=h$, while $\Tpol^{ij}_{hh}=\TTpol^{ij}_{hh}=0$; the tensors select the
helicity-flip channel with $|h'-h|=2$. We parametrize the calculated
time-even overlap with eight GTMDs. Their common arguments
$(x,k_\perp^2,\kb\cdot\Db,\Delta_\perp^2)$ are suppressed:
\begin{align}
 W_{h'h}^{[\gamma^+]}
 &=\Ipol_{h'h}\FS
 +i\Vpol_{h'h}\frac{\kb\times\Db}{m^2}\FA
 \nonumber\\[-2pt]
 &\quad+\frac{\Tpol^{ij}_{h'h}k_\perp^ik_\perp^j}{m^2}\FK
 +\frac{\Tpol^{ij}_{h'h}\Delta_\perp^i\Delta_\perp^j}{m^2}\FD,
 \label{eq:vector}
\end{align}
\begin{align}
 W_{h'h}^{[\gamma^+\gamma_5]}
 &=\Vpol_{h'h}\GS
 +i\Ipol_{h'h}\frac{\kb\times\Db}{m^2}\GA
 \nonumber\\[-2pt]
 &\quad+\frac{\TTpol^{ij}_{h'h}k_\perp^ik_\perp^j}{m^2}\GK
 +\frac{\TTpol^{ij}_{h'h}\Delta_\perp^i\Delta_\perp^j}{m^2}\GD .
 \label{eq:axial}
\end{align}
Here $\mathcal F_X$ and $\mathcal G_X$ denote the vector ($\gamma^+$)
and axial ($\gamma^+\gamma_5$) GTMDs, respectively, with
$X\in\{S,A,Dk,D\Delta\}$.
For $M_\gamma=0$, we normalize tensor prefactors with the fermion mass $m$.
Using a reference scale $\mu$ in these prefactors rescales the corresponding GTMDs by
$\mu^2/m^2$ and requires $1/\mu^2$ in their moment definitions, preserving
the correlator and observables.

Matching the photon-helicity overlaps gives
\begin{align}
 \FS&=\Cf\frac{m^2+c\,\boldsymbol q_+\cdot\boldsymbol q_-}{D_+D_-},
 &\FA&=-\Cf\frac{c\beta m^2}{D_+D_-},\nonumber\\
 \FK&=-2\Cf\frac{x\beta m^2}{D_+D_-},
 &\FD&=\Cf\frac{x\beta^3m^2}{2D_+D_-},
 \label{eq:Fresults}\\
 \GS&=\Cf\frac{m^2+d\,\boldsymbol q_+\cdot\boldsymbol q_-}{D_+D_-},
 &\GA&=-\Cf\frac{d\beta m^2}{D_+D_-},\nonumber\\
 \GK&=\GD=0.
 \label{eq:Gresults}
\end{align}
Here $c=x^2+\beta^2$ and $d=2x-1$; the denominators and normalization
factors are
\begin{equation}
 D_\pm=\boldsymbol q_\pm^{\,2}+m^2,\quad
 \Cf=\frac{N_c\alpha e_f^2}{2\pi^2},\quad \af=\pi\Cf.
 \label{eq:definitions}
\end{equation}
The factor $\af$ normalizes the distributions after transverse integration.
The axial photon-helicity-flip GTMDs $\GK$ and $\GD$ vanish in the
leading two-particle calculation. Thus six of the eight GTMDs are nonzero.

\begin{figure*}[!t]
 \centering
\includegraphics[width=0.92\textwidth]{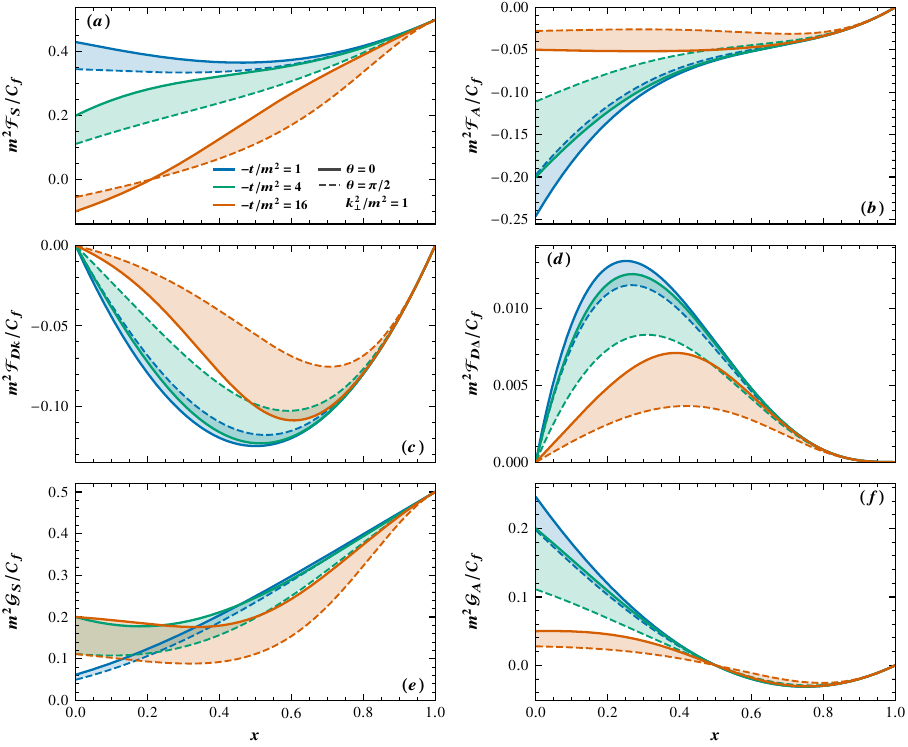}
 \caption{All six nonzero photon GTMDs in units of $\Cf/m^2$ at
 $k_\perp^2/m^2=1$: (a) $\FS$, (b) $\FA$, (c) $\FK$, (d) $\FD$,
 (e) $\GS$, and (f) $\GA$. Colors identify $-t/m^2=1$, $4$, and $16$; for
 each transfer the solid and dashed curves are
$\theta=\angle(\kb,\Db)=0$ and $\pi/2$, and the band contains all
intermediate angles.}
 \label{fig:gtmd}
\end{figure*}

Figure~\ref{fig:gtmd} shows their $x$ dependence at
$k_\perp^2/m^2=1$. The transfer enters through $\beta\Db$, so its effect
vanishes as $x\to1$. At $-t\lesssim4m^2$, $\FA$ and $\GA$ have peak
magnitudes comparable to $\FS$ and $\GS$; their relative size decreases
at larger transfer. The peak of $\FD$ is more than an order of magnitude
smaller than those of $\FS$ and $\GS$.
The factor $d=2x-1$ fixes the zero of $\GA$.

\paragraph*{Projection limits.}

In the forward limit, $\Db=0$, the GTMDs $\FS$ and $\GS$ reduce to
the unpolarized TMD $f_1^\gamma(x,k_\perp^2)$ and helicity TMD
$g_{1L}^\gamma(x,k_\perp^2)$,
respectively:
\begin{equation}
 f_1^\gamma=\Cf\frac{m^2+c\,k_\perp^2}{(k_\perp^2+m^2)^2},\qquad
 g_{1L}^\gamma=\Cf\frac{m^2+d\,k_\perp^2}{(k_\perp^2+m^2)^2}.
 \label{eq:tmds}
\end{equation}
These reproduce the basis light-front quantization benchmarks
\cite{Nair:2022evk,Nair:2023lir} and the real-photon results of
Ref.~\cite{puhan2025photontmd}.
At $\Db=0$, the helicity-flip amplitude is
$W_{+-}^{[\gamma^+]}=-k_L^2\FK/m^2$, with $k_L=k_x-i k_y$. In the
fermion-mass normalization of Eq.~\eqref{eq:vector}, we define
$f_{1TT}^\gamma(x,k_\perp^2)\equiv\FK|_{\Db=0}$,
the real-photon counterpart of the tensor-polarized coefficient $f_{1TT}$
\cite{bacchetta2000spinone}.
Reference~\cite{puhan2025photontmd} treats real and virtual photons and
normalizes $f_{1TT}$ with $M_\gamma^2$. Our definition fixes the
coefficient directly from the physical-helicity overlap of a
real photon.

Integrating $\FS$ and $\GS$ over $\kb$ at fixed momentum transfer
gives the helicity-nonflip vector and axial GPDs, respectively, at zero
skewness:
\begin{align}
 F^f(x,t)
 &=\af\left[
 c\log\frac{\Luv^2}{m^2}
 +2\beta(x-c\beta\tau)\mathcal S(\beta\sqrt\tau)
 \right],\nonumber\\
 \widetilde F^f(x,t)
 &=\af\left[
 d\log\frac{\Luv^2}{m^2}
 +2\beta(1-d\beta\tau)\mathcal S(\beta\sqrt\tau)
 \right].
 \label{eq:gpds}
\end{align}
The GPDs are $F^f=H^{[\gamma^+]}_{hh}$ and
$\widetilde F^f=h\,H^{[\gamma^+\gamma_5]}_{hh}$, where
$H^{[\Gamma]}_{h'h}(x,0,t)=\int\dd^2k_\perp W^{[\Gamma]}_{h'h}$.
The integration uses a shifted circular cutoff $\Luv$ (SM).
Here $\tau=-t/(4m^2)$ and
$\mathcal S(z)=\operatorname{arsinh}z/[z\sqrt{1+z^2}]$, with $\mathcal S(0)=1$.
The terms multiplying $\FA$ and $\GA$ integrate to zero by reflection
of the momentum component perpendicular to $\Db$.
The logarithms reproduce the nonflip photon-GPD kernels
\cite{mukherjee2011generalized}. The finite terms use the on-shell
denominator and stated cutoff; $O((m^2-t)/\Luv^2)$ terms are omitted. For
$\Db=\Delta_\perp\hat{\boldsymbol x}$, transverse integration of the
$\FK$ and $\FD$ structures gives the double-helicity-flip GPD
\begin{align}
 E_1(x,t)
 &=-\frac{1}{2m^2}\!\int\!\dd^2k_\perp
 \left[(k_x^2-k_y^2)\FK+\Delta_\perp^2\FD\right]\nonumber\\
 &=-\af x\beta\left[1-\mathcal S(\beta\sqrt\tau)\right],
 \label{eq:E1}
\end{align}
with $H_{+-}^{[\gamma^+]}=2E_1$ in our helicity convention.
Its quadrupole structure agrees with Ref.~\cite{mukherjee2013generalized}.
At $t=0$, these GPDs give the unpolarized and helicity PDFs; their
leading logarithms (LL) are
\begin{equation}
 q^\gamma_{\rm LL}=\af c\log\frac{\Luv^2}{m^2},\qquad
 \Delta q^\gamma_{\rm LL}=\af d\log\frac{\Luv^2}{m^2},
 \label{eq:pdfs}
\end{equation}
where $q^\gamma$ and $\Delta q^\gamma$ are the unpolarized and helicity
distributions. The coefficients are the leading-order photon-to-fermion splitting kernels
\cite{Nisius2000photon,StratmannVogelsang1996photon}.

\begin{table}[!tb]
 \footnotesize
 \setlength{\tabcolsep}{2.2pt}
 \caption{TMD, GPD, PDF, and GFF projections of the six nonzero GTMDs in
 Eqs.~\eqref{eq:Fresults}--\eqref{eq:Gresults}, together with their orbital
 and spin--orbit moments, at leading two-particle order. A dash denotes no
 contribution to a projection.
 Daggers mark GTMDs that contribute jointly to the same GPD or GFF.
 The last column gives $k_\perp^2$ moments; $F_1-1$ uses the forward
 subtraction in Eq.~\eqref{eq:F1sum}.}
 \label{tab:limits}
 \begin{ruledtabular}
 \begin{tabular}{cccccc}
 GTMD & TMD & GPD & PDF & GFF & $k_\perp^2$ mom.\\
 \hline
 $\FS$ & $f_1^\gamma$ & $F^f$ & $q^\gamma$ & $F_1-1$ & --\\
 $\FA$ & -- & -- & -- & -- & $\ell_z^f$\\
 $\FK$ & $f_{1TT}^\gamma$ & $E_1^\dagger$ & -- & $(2tF_3)^\dagger$ & --\\
 $\FD$ & -- & $E_1^\dagger$ & -- & $(2tF_3)^\dagger$ & --\\
 $\GS$ & $g_{1L}^\gamma$ & $\widetilde F^f$ & $\Delta q^\gamma$ & -- & --\\
 $\GA$ & -- & -- & -- & -- & $\mathcal C_z^f$
 \end{tabular}
 \end{ruledtabular}
\end{table}

Table~\ref{tab:limits} collects the projections. The $\FD$ contribution
vanishes at $\Db=0$; $\FK$ gives $f_{1TT}^\gamma$ but has zero angular
integral in a PDF. The orbital information resides in the weighted
$\FA$ and $\GA$ moments.

\paragraph*{Orbital motion and spin--orbit correlation.}

\begin{figure}[!htb]
 \centering
 \includegraphics[width=\linewidth]{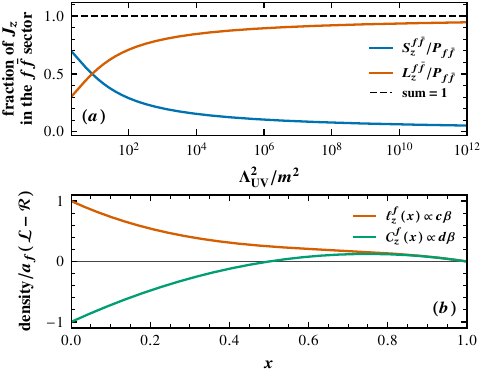}
\caption{For photon helicity $+1$: (a) total constituent-spin and orbital
 fractions of $J_z$ in the leading fixed-flavor $f\bar f$ sector,
 Eq.~\eqref{eq:fractions}. (b) Single-fermion orbital and spin--orbit
 densities divided by $\af(\Lcut-\Rcut)$, giving $c\beta$ and $d\beta$,
 respectively; $\Lcut$ and $\Rcut$ are defined in Eq.~\eqref{eq:LR}.}
 \label{fig:angmom}
\end{figure}

We define the fermion OAM $\ell_z^f$ and spin--orbit correlation
$\mathcal C_z^f$ using light-front canonical (Jaffe--Manohar) operators
with ordinary transverse derivatives \cite{jaffe1990spin,lorce2014spinorbit}.
The GTMDs $\FA$ and $\GA$ determine these moments, respectively.
For equal initial and final photon helicities,
$h'=h$, $\FA$ multiplies $ih\,\kb\times\Db/m^2$ in the vector correlator,
while $\GA$ multiplies $i\,\kb\times\Db/m^2$ in the axial correlator.
Their moments follow the construction for the nucleon GTMDs
$F_{1,4}$ and $G_{1,1}$
\cite{lorce2011overlap,lorce2012oam,lorce2014spinorbit}.
With our signs, the corresponding coefficients are $\FA$ and $-\GA$.
For photon helicity $h=+1$,
\begin{equation}
 (\ell_z^f,\mathcal C_z^f)
 =-\int_0^1\!\dd x\int\!\dd^2k_\perp\,
 \frac{k_\perp^2}{m^2}\,(\FA,\GA),
\label{eq:oamdef}
\end{equation}
both evaluated at $\Db=\boldsymbol0_\perp$. The OAM adds the orbital
contributions from the two fermion helicities $s=\pm1$. The spin--orbit
correlation measures the alignment of the fermion spin with its orbital
motion. The axial insertion weights each orbital contribution by $s$,
giving the $s=+1$ contribution minus the $s=-1$ contribution.

Both moments depend logarithmically on the ultraviolet regulator and its
scheme. We use a common, $x$-independent disk $k_\perp^2<\Luv^2$ and write
\begin{equation}
 \Lcut=\log\!\left(1+\frac{\Luv^2}{m^2}\right),
 \qquad
 \Rcut=\frac{\Luv^2}{\Luv^2+m^2}.
 \label{eq:LR}
\end{equation}
At $\Db=0$, $D_+D_-=(k_\perp^2+m^2)^2$, giving
\begin{equation}
 (\ell_z^f,\mathcal C_z^f)
 =\af(\Lcut-\Rcut)\!\int_0^1\!\dd x\,\beta(c,d).
 \label{eq:factorization}
\end{equation}
To understand the sign and
relative size of the two moments, it is instructive to separate the
helicity-channel contributions already summed in $c$ and $d$.
The factor $\beta=1-x$ is the active fermion's share of the pair's relative
OAM about the transverse center of momentum. For $h=+1$, the two
OAM-carrying channels are $(s,\bar s)=(+,-)$ and $(-,+)$, both with
$\ell_z=+1$ and with the active-fermion spin aligned and anti-aligned,
respectively. Their wave-function numerators are proportional to
$x(q_x+iq_y)$ and $-\beta(q_x+iq_y)$, respectively (SM). Both channels
have the same denominator and normalization, so taking their absolute
squares gives longitudinal weights $x^2$ and $\beta^2$. Multiplying these
weights by the fermion's orbital share $\beta$ gives the channel
contributions $x^2\beta$ and $\beta^3$:
\begin{equation*}
 \begin{aligned}
 \int_0^1x^2\beta\,\dd x&=\frac1{12}, &
 \int_0^1\beta^3\,\dd x&=\frac14 .
 \end{aligned}
 \end{equation*}
The anti-aligned channel therefore has three times the integrated weight of
the aligned channel. Their sum and signed difference give
$\int_0^1c\beta\,\dd x=1/3$ and
$\int_0^1d\beta\,\dd x=-1/6$, respectively, so that
\begin{equation}
 \boxed{\;
 \ell_z^f=\frac{\af}{3}(\Lcut-\Rcut),\qquad
 \mathcal C_z^f=-\frac{\ell_z^f}{2}. \;}
 \label{eq:oamresult}
\end{equation}
The common transverse integral cancels in the ratio for any
$x$-independent transverse regulator.
Reversing $h,s,\bar s$ preserves the channel weights. Since
$\ell_z=h-(s+\bar s)/2$, this gives
$\ell_z^f(-h)=-\ell_z^f(h)$, while the spin--orbit weight $s\ell_z$ is
unchanged. Thus $\mathcal C_z^f(h)=-h\,\ell_z^f(h)/2$ remains negative
for both photon helicities. For an unpolarized photon, the OAM averages
to zero while the spin--orbit correlation remains negative.

In the perturbative dressed-quark model, all OAM-carrying amplitudes have the
same active-quark helicity, so
$\mathcal C_z^q=\ell_z^q$
\cite{mukherjee2014wigner}; the photon has two OAM-carrying channels with
opposite active-fermion helicity.

Figure~\ref{fig:angmom}(b) shows the orbital and spin--orbit densities
$\ell_z^f(x)$ and $\mathcal C_z^f(x)$, defined by the integrands of the
$x$ integral in Eq.~\eqref{eq:oamdef}, divided by $\af(\Lcut-\Rcut)$.
At $x\to0$, the anti-aligned channel dominates, giving normalized
densities $+1$ and $-1$. Equal channel weights make
$\mathcal C_z^f(x)$ vanish at $x=1/2$, while the OAM remains positive.
The aligned channel makes the spin--orbit density positive for $x>1/2$.
Both densities vanish as the fermion's orbital share $\beta\to0$.
Their ratio is $\mathcal C_z^f(x)/\ell_z^f(x)=d/c$ for a common
spin-independent transverse weight.
Charge conjugation gives $\mathcal C_z^{\bar f}=\mathcal C_z^f$.

We next examine how constituent spin and orbital motion share the angular
momentum of the photon's two-particle sector, as illustrated in
Fig.~\ref{fig:angmom}(a).
For $h=+1$, the bare pair probability and total constituent spin are
\begin{equation*}
 \begin{aligned}
 P_{f\bar f}&=N_c\sum_{s,\bar s}\int_0^1\!\dd x\int\!\dd^2q_\perp\,
 |\Psi^+_{s\bar s}(x,\qb)|^2,\\
 S_z^{f\bar f}&=N_c\sum_{s,\bar s}\int_0^1\!\dd x\int\!\dd^2q_\perp\,
 \frac{s+\bar s}{2}|\Psi^+_{s\bar s}(x,\qb)|^2.
 \end{aligned}
\end{equation*}
Here $(s+\bar s)/2$ is the total constituent helicity.
Inserting Eq.~\eqref{eq:wavefunction} and integrating
with the same transverse cutoff gives
\begin{equation}
 S_z^{f\bar f}=\af\Rcut,
 \qquad
 P_{f\bar f}=\af\left(\tfrac23\Lcut+\tfrac13\Rcut\right).
 \label{eq:spinprob}
\end{equation}
The state normalization gives
$Z_\gamma^{-1}=1+P_{f\bar f}$ and $Z_\gamma=1-P_{f\bar f}+O(\alpha^2)$.
The perturbative truncation requires $P_{f\bar f}\ll1$.
The spin $S_z^{f\bar f}$ includes both constituents. Charge conjugation
gives equal integrated fermion and antifermion OAM, so
$L_z^{f\bar f}=2\ell_z^f$ and
\begin{equation}
 Z_\gamma+S_z^{f\bar f}+L_z^{f\bar f}=1+O(\alpha^2).
 \label{eq:budget}
\end{equation}
Since $J_z=h$ holds channel by channel, this checks that the $\FA$ moment
reproduces the wave-function OAM.
The leading logarithm of the pair OAM and this helicity sum rule have
analogs for a polarized gluon \cite{harindranath1999orbital};
here the GTMDs also resolve the fermion share and spin--orbit correlation.

Figure~\ref{fig:angmom}(a) plots, within the regulated bare two-body sector,
the fractions of angular momentum carried by spin and orbital motion,
\begin{equation}
 \frac{S_z^{f\bar f}}{P_{f\bar f}}
 =\frac{3\Rcut}{2\Lcut+\Rcut},\qquad
 \frac{L_z^{f\bar f}}{P_{f\bar f}}
 =\frac{2(\Lcut-\Rcut)}{2\Lcut+\Rcut},
 \label{eq:fractions}
\end{equation}
so the two curves sum to unity at every cutoff.

At large cutoff, the spin approaches $\af$ while the OAM grows
logarithmically, making the orbital fraction approach unity.

\paragraph*{GTMD sum rules and gravitational form factors.}

\begin{figure}[!t]
 \centering
 \includegraphics[width=\linewidth]{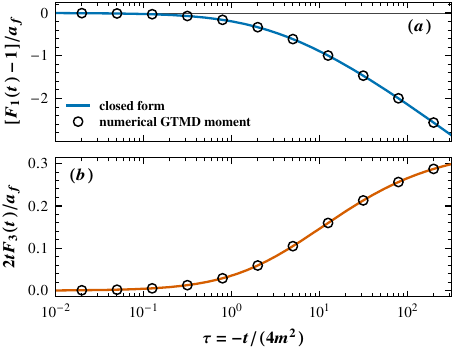}
 \caption{Photon GFFs (a) $(F_1-1)/\af$ and (b) $2tF_3/\af$.
 Solid curves: analytic results, Eqs.~\eqref{eq:F1closed}--\eqref{eq:F3closed};
 open symbols: numerical GTMD sum rules, Eqs.~\eqref{eq:F1sum}--\eqref{eq:F3sum}.}
 \label{fig:gff}
\end{figure}

The $x$-weighted (second Mellin) moment of the vector bilocal gives the
fermion energy-momentum tensor (EMT) matrix element
\cite{ji1997gpd,polyakov2018gff}.
We normalize the full $T^{++}$ matrix element by $2(P^+)^2$. In Milton's
photon GFF basis \cite{milton1977photon}, its diagonal entry is $F_1(t)$
and helicity flip determines $F_3(t)$. Adding the equal antifermion
contribution and the one-photon normalization term gives the GTMD sum rules
for $F_1$ and $F_3$, up to $O(\alpha^2)$ corrections:
\begin{align}
 F_1(t)-1
 &=2\int_0^1\!\dd x\,x\!\int\!\dd^2k_\perp\nonumber\\[-2pt]
 &\quad\times
 \left[\FS(x,\kb,\Db)-\FS(x,\kb,\boldsymbol0_\perp)\right],
 \label{eq:F1sum}\\
 2tF_3(t)
 &=\frac{2}{m^2}\int_0^1\!\dd x\,x\!\int\!\dd^2k_\perp
 \nonumber\\
 &\quad\times\big[(k_x^2-k_y^2)\FK+\Delta_\perp^2\FD\big],
 \label{eq:F3sum}
\end{align}
for $\Db=\Delta_\perp\hat{\boldsymbol x}$. The one-photon normalization
$Z_\gamma=1-P_{f\bar f}+O(\alpha^2)$ supplies the forward subtraction in
Eq.~\eqref{eq:F1sum}, removing the ultraviolet logarithm and enforcing
$F_1(0)=1$. Comparison with Eq.~\eqref{eq:E1} gives
$2tF_3=-4\int_0^1\dd x\,x\,E_1(x,t)$, so the quadrupole that defines $E_1$
also fixes $F_3$. The resulting closed forms are
\begin{align}
 \frac{F_1(t)-1}{\af}
 &=\frac{35}{18}-\frac{13}{6\tau}
 -\frac{(4\tau-5)r\eta}{3\tau}
 -\frac{(2\tau-1)\eta^2}{2\tau^2},
 \label{eq:F1closed}\\
 \frac{2tF_3(t)}{\af}
 &=\frac13-\frac7\tau+\frac{6r\eta}{\tau}
 -\frac{(2\tau-1)\eta^2}{\tau^2}.
 \label{eq:F3closed}
\end{align}
Here $\eta=\operatorname{arsinh}\sqrt\tau$ and $r=\sqrt{1+1/\tau}$.
For a unit-charge lepton, $\af=\alpha/(2\pi)$, and these expressions agree
term by term with the one-loop QED results of Ref.~\cite{freese2022spatial},
which reproduce Refs.~\cite{berends1976photon,milton1977photon}.
Figure~\ref{fig:gff} compares them with direct numerical integration of
Eqs.~\eqref{eq:F1sum}--\eqref{eq:F3sum} (SM).

The remaining form factor $F_2$ probes the transverse EMT. For
$\Db=Q\hat{\boldsymbol x}$, $Q=\sqrt{-t}$, the full matrix element in
Milton's basis \cite{milton1977photon}, with our helicity conventions, obeys
$\langle p',+|T^{yy}|p,-\rangle=-t^2F_2(t)$ and
$\langle p',+|T^{xx}|p,-\rangle=0$.
Subtracting the zero-transfer pair overlap at the same transverse cutoff
before removing the cutoff gives (SM)
\begin{equation}
 \frac{2tF_2(t)}{\af}
 =\frac13+\frac3\tau-\frac{2r\eta}{\tau}
 -\frac{\eta^2}{\tau^2},
 \label{eq:F2closed}
\end{equation}
which matches the one-loop QED result in Eq.~(35c) of
Ref.~\cite{freese2022spatial}.

\paragraph*{Conclusion.}

We have calculated the leading-twist, time-even, chiral-even fermion
GTMDs of a real photon at zero skewness and order $\alpha$.
Six of the eight GTMDs are nonzero. The $\FA$ and $\GA$ moments trace
canonical OAM and spin--orbit correlation to two orbital channels
with opposite fermion helicities. Their $1{:}3$ aligned-to-anti-aligned
weights explain $\mathcal C_z^f=-\ell_z^f/2$ for photon helicity $+1$.
The negative sign reflects the greater weight of the anti-aligned channel.
In the dressed-quark model, a single active-quark helicity instead
gives $\mathcal C_z^q=\ell_z^q$.
The ratio $\mathcal C_z^f/\ell_z^f=-1/2$ is exact for a common
$x$-independent transverse regulator,
although each moment is cutoff dependent.

The pair's OAM contains the ultraviolet logarithm, while its spin remains
finite; adding the bare photon satisfies the angular-momentum sum rule.
Agreement with photon TMDs, splitting kernels, and all three one-loop
GFFs tests the same wave function through parton projections and distinct
EMT components: $F_1$ and $F_3$ follow from GTMD sum rules, while $F_2$
requires the transverse EMT. These closed forms provide a perturbative reference for
orbital and spin--orbit structure. Higher Fock sectors and radiative
corrections can test how additional helicity channels modify the ratio.

\FloatBarrier
\paragraph*{Acknowledgments.}

S.N.\ acknowledges support from the Fundamental Research Funds for the
Central Universities under Grant Nos.~lzujbky-2023-stlt01,
lzujbky-2024-oy02, and lzujbky-2025-eyt01, and from the National Natural
Science Foundation of China under Grant No.~12335003.
\textit{AI use:} OpenAI's GPT-6 Astra and Anthropic's Claude Opus 5.5 were
used for proofreading and assistance with two rounds of simulated peer
review.

\vspace*{-6pt}
 \bibtolerant
\bibliographystyle{apsrev4-2}
\bibliography{references}

\onecolumngrid
\clearpage
\begin{center}
 {\large\bfseries Supplemental Material for\\[3pt]
 ``\PaperTitle''}
\end{center}
\twocolumngrid
\setcounter{section}{0}
\setcounter{subsection}{0}
\setcounter{equation}{0}
\setcounter{figure}{0}
\setcounter{table}{0}
\renewcommand{\thesection}{S\arabic{section}}
\renewcommand{\thesubsection}{S\arabic{section}.\arabic{subsection}}
\renewcommand{\theequation}{S\arabic{equation}}
\renewcommand{\thefigure}{S\arabic{figure}}
\renewcommand{\thetable}{S\arabic{table}}

\section{Helicity overlaps and polarization structure}
\label{supp:overlap}
\raggedbottom

\subsection{Helicity overlaps}
\label{supp:convention}

The two-particle wave function in Letter Eq.~\Leqref{eq:wavefunction}
has the helicity numerators
\begin{equation}
 \begin{aligned}
 N^+&=\sqrt2\begin{pmatrix}im&xq_R\\-\beta q_R&0\end{pmatrix},\\
 N^-&=\sqrt2\begin{pmatrix}0&\beta q_L\\-xq_L&-im\end{pmatrix}.
 \end{aligned}
 \label{eqS:amplitudes}
\end{equation}
Here $N^h_{s\bar s}=\chi_s^\dagger \mathcal K_h\chi_{-\bar s}$,
$\beta=1-x$, and $q_{R,L}=q_x\pm iq_y$.
The fermion and antifermion light-front helicities $s/2$ and $\bar s/2$
label the rows and columns, respectively, with $s,\bar s=\pm1$;
$h=\pm1$ is the photon helicity.
The mass terms have $\ell_z=0$ and the momentum terms $\ell_z=h$.
The $f\bar f$ component is generated by annihilating the photon in the
initial ket, so the vertex selects the term
$\epsilon_\perp^{h\,i}a_h$ in $A_\perp^i$ and the wave function contains
the unstarred $\epsilon_\perp^h$
\cite{lappi2020relativistic,beuf2016dipole}.
The conjugate vector multiplies the creation operator $a_h^\dagger$,
as in the field expansion of Freese and Cosyn
[Eq.~(10) of Ref.~\cite{freese2022spatial}].
The starred form printed in
Refs.~\cite{harindranath1999radiative,mukherjee2011generalized,mukherjee2013generalized}
uses $\epsilon_\perp^{h*}=-\epsilon_\perp^{-h}$ in this circular basis,
so matching its numerator requires the helicity relabeling $h\to-h$
and an overall phase.
For a real photon the parent invariant mass vanishes. The vertex and
light-front energy denominator contain the same longitudinal factor:
\begin{equation}
 \frac{\mathcal K_h/[x(1-x)]}
 {-(q_\perp^2+m^2)/[x(1-x)]}
 =-\frac{\mathcal K_h}{q_\perp^2+m^2}.
 \label{eqS:energydenominator}
\end{equation}
This cancellation gives the denominator in Letter
Eq.~\Leqref{eq:wavefunction}; $m$ remains the constituent mass.

For $h=+1$, the two OAM-carrying entries of Eq.~\eqref{eqS:amplitudes}
give
\begin{equation*}
 \begin{aligned}
 |N^+_{+,-}|^2&=|\sqrt2\,xq_R|^2=2x^2q_\perp^2,\\
 |N^+_{-,+}|^2&=|-\sqrt2\,\beta q_R|^2=2\beta^2q_\perp^2.
 \end{aligned}
\end{equation*}
Here $|q_R|^2=q_\perp^2$. The two wave functions share the same
denominator and normalization, leaving longitudinal weights $x^2$ and
$\beta^2$. The minus sign in $N^+_{-,+}$ drops out of its absolute square;
the relative minus sign in the spin--orbit moment instead comes from
the axial weight $s$. Including the mass term, the channel sums are
\begin{equation}
 \sum_{s\bar s}\!\!\begin{pmatrix}1\\ (s+\bar s)/2\\ \ell_z\\ s\,\ell_z
 \end{pmatrix}\!\!|N^+_{s\bar s}|^2
 =2\begin{pmatrix}m^2+cq_\perp^2\\ m^2\\ cq_\perp^2\\ d\,q_\perp^2
 \end{pmatrix}.
 \label{eqS:weights}
\end{equation}
Here $c=x^2+\beta^2$ and $d=2x-1$.
Multiplication by $\Cf/[2(q_\perp^2+m^2)^2]$ and integration give the pair
probability $P_{f\bar f}$, spin $S_z^{f\bar f}$, and orbital angular momentum
$L_z^{f\bar f}$; the fourth entry acquires the fermion orbital share
$\beta$ for $\mathcal C_z^f$.
Here $\Cf=N_c\alpha e_f^2/(2\pi^2)$, $\af=\pi\Cf$,
$\alpha=e^2/(4\pi)$, and $N_c=1$ for a lepton and $3$ for a quark flavor.

At zero skewness, spinor completeness gives the helicity correlators
\begin{equation}
 W_{h'h}^{[\Gamma]}
 =\frac{\Cf}{2D_+D_-}
 \operatorname{tr}\left[
 S^{[\Gamma]} \mathcal K_h(\boldsymbol q_-)
 \mathcal K_{h'}^\dagger(\boldsymbol q_+)\right].
 \label{eqS:traces}
\end{equation}
Here $\Db$ is the transverse momentum transfer,
$\boldsymbol q_\pm=\kb\pm\beta\Db/2$ are the relative transverse momenta
in the final ($+$) and initial ($-$) wave functions, and
$D_\pm=\boldsymbol q_\pm^{\,2}+m^2$;
$S^{[\gamma^+]}=\id$ and $S^{[\gamma^+\gamma_5]}=\sigma^3$.
The label $h'$ denotes the final photon helicity.
The helicity projections needed below are
\begin{equation}
 \begin{aligned}
 W_{hh}^{[\gamma^+]}&=\FS+ih\frac{\kb\times\Db}{m^2}\FA,\\
 W_{hh}^{[\gamma^+\gamma_5]}&=h\,\GS+i\frac{\kb\times\Db}{m^2}\GA,\\
 W_{+-}^{[\gamma^+]}
 &=-\frac{k_L^2\FK+\Delta_L^2\FD}{m^2},
 \end{aligned}
 \label{eqS:helicity}
\end{equation}
where $k_L=k_x-ik_y$, $\Delta_L=\Delta_x-i\Delta_y$, and
$W_{-+}^{[\gamma^+]}=(W_{+-}^{[\gamma^+]})^*$, with
$W_{+-}^{[\gamma^+\gamma_5]}=W_{-+}^{[\gamma^+\gamma_5]}=0$.
The two opposite-fermion-helicity channels have equal helicity-flip
products. Their opposite axial weights $s=\pm1$ cancel, giving
$\GK=\GD=0$ for this two-body overlap.

\subsection{Polarization decomposition and tensor redundancy}
\label{supp:basis}

The transverse polarization matrix decomposes as
\begin{equation}
 B^{ij}_{h'h}=\frac12\delta^{ij}\Ipol_{h'h}
 +\frac{i}{2}\varepsilon_\perp^{ij}\Vpol_{h'h}
 +\frac12\Tpol^{ij}_{h'h}.
 \label{eqS:poldecomposition}
\end{equation}
Here $B^{ij}_{h'h}=\epsilon_\perp^{h'\,*i}\epsilon_\perp^{h\,j}$,
$\Ipol_{h'h}=B^{ii}_{h'h}$, $\Vpol_{h'h}=-i\varepsilon_\perp^{ij}B^{ij}_{h'h}$, and
$\Tpol^{ij}_{h'h}=B^{ij}_{h'h}+B^{ji}_{h'h}-\delta^{ij}\Ipol_{h'h}$, with
$\varepsilon_\perp^{12}=1$. A $2\times2$ matrix has one trace component,
one antisymmetric component, and two symmetric traceless components.
These exhaust the transverse polarization structures. Under transverse
rotations, $\Ipol_{h'h}$ and $\Vpol_{h'h}$ are scalars; $\Vpol_{h'h}$ changes sign under
a transverse reflection and represents longitudinal vector polarization
(helicity). The tensor components describe linear polarization.

For a diagonal polarization state, $B^{ji}_{hh}=(B^{ij}_{hh})^*$, so
$\varepsilon_\perp^{ij}B^{ij}_{hh}$ is purely imaginary.
The factor $-i$ makes $\Vpol_{hh}$ real and gives $\Vpol_{hh}=h$ for
$\boldsymbol\epsilon_\perp^h=(-h,-i)/\sqrt2$.
The dual tensor and its helicity projections are
\begin{equation}
 \begin{gathered}
 \TTpol^{ij}_{h'h}=i\varepsilon_\perp^{ik}\Tpol^{kj}_{h'h},\\
 \Ipol_{h'h}=\delta_{h'h},\quad \Vpol_{h'h}=h\delta_{h'h},\\
 \Tpol_{hh}=\TTpol_{hh}=0,\qquad
 \Tpol_{+-}=\begin{pmatrix}-1&i\\ i&1\end{pmatrix},\\
 \Tpol_{-+}=\Tpol_{+-}^*,\qquad
 \TTpol_{+-}=\Tpol_{+-},\quad \TTpol_{-+}=-\Tpol_{-+}.
 \end{gathered}
 \label{eqS:polhelicities}
\end{equation}
The subscripts denote final and initial photon helicities; transverse
matrix indices are suppressed in the last two lines.
Contracting with $\varepsilon_\perp^{ik}$ preserves symmetry and
tracelessness in two dimensions and gives the opposite reflection parity
needed in the axial correlator. The factor $i$ fixes the displayed helicity
phases.

The symmetric traceless transverse contractions obey
\begin{equation}
 \Delta_\perp^2 U_{kk}-2\omega U_{k\Delta}
 +k_\perp^2 U_{\Delta\Delta}=0.
 \label{eqS:tensoridentity}
\end{equation}
Here $U_{ab}=U^{ij}a_\perp^i b_\perp^j$, with $U^{ij}$ symmetric
and traceless, and $\omega=\kb\cdot\Db$.
The mixed contraction can be eliminated for $\omega\ne0$.
Extracting $\FK$ and $\FD$ from $-m^2W_{+-}^{[\gamma^+]}$ has determinant
$-2\omega(\kb\times\Db)$. For the calculated overlap, the explicit
GTMDs have positive denominators $D_+D_-$ for $m>0$ and no pole at
$\omega=0$. Their forward limits are
\begin{equation}
 \begin{aligned}
 \FK\big|_{\Db=0}
 &=-\frac{2\Cf x\beta m^2}{(k_\perp^2+m^2)^2},\\
 \FD\big|_{\Db=0}
 &=\frac{\Cf x\beta^3m^2}{2(k_\perp^2+m^2)^2}.
 \end{aligned}
 \label{eqS:smoothlimits}
\end{equation}
These functions extend smoothly to parallel and orthogonal momenta.
At $\Db=0$, the tensor multiplying $\FD$ vanishes; the coefficient in
Eq.~\eqref{eqS:smoothlimits} is determined by continuity from nonzero $\Db$.
The eight-function parametrization is specific to the calculated
time-even overlap.

\section{Transverse integrals}
\label{supp:integrals}

The common denominator is
\begin{equation}
 D_+D_-=
 \left(m^2+k_\perp^2+\frac{\beta^2\Delta_\perp^2}{4}\right)^2
 -\beta^2(\kb\cdot\Db)^2 .
 \label{eqS:denominator}
\end{equation}
Feynman parametrization gives
\begin{align}
 \frac1{D_+D_-}
 &=\int_0^1\frac{\dd u}
 {[\kappa_\perp^2+\mu_u^2]^2},\nonumber\\
 \mu_u^2&=m^2+u(1-u)\beta^2\Delta_\perp^2 .
 \label{eqS:feynman}
\end{align}
Here $\boldsymbol\kappa_\perp=\kb+(2u-1)\beta\Db/2$ is the shifted
integration momentum. The full-plane integral is
$\int\dd^2\kappa_\perp/(\kappa_\perp^2+\mu_u^2)^2=\pi/\mu_u^2$.
With the shifted circular cutoff $\kappa_\perp^2<\Luv^2$,
\begin{align}
 \int_{\kappa_\perp^2<\Luv^2}
 \frac{\dd^2\kappa_\perp\,\kappa_\perp^2}
 {[\kappa_\perp^2+\mu_u^2]^2}
 &=\pi\left[
 \log\left(1+\frac{\Luv^2}{\mu_u^2}\right)\right.\nonumber\\
 &\qquad\left.-\frac{\Luv^2}{\Luv^2+\mu_u^2}\right].
 \label{eqS:radial}
\end{align}
The parameter integrals give
\begin{align}
 \int_0^1\frac{\dd u}{\mu_u^2}
 &=\frac{\mathcal S(\beta\sqrt\tau)}{m^2},\nonumber\\
 \beta^2\Delta_\perp^2
 \int_0^1\frac{u(1-u)\dd u}{\mu_u^2}
 &=1-\mathcal S(\beta\sqrt\tau).
 \label{eqS:parametric}
\end{align}
Here the transfer variable is $\tau=-t/(4m^2)$, with
$t=-\Delta_\perp^2$, and the function is
$\mathcal S(z)=\operatorname{arsinh}z/[z\sqrt{1+z^2}]$, with
$\mathcal S(0)=1$.
The exact finite-cutoff parton densities are
\begin{equation}
 \bigl(q^\gamma,\Delta q^\gamma\bigr)(x,\Luv^2)
 =\af\bigl[(c,d)\Lcut+2\beta(x,1)\Rcut\bigr].
 \label{eqS:pdfs}
\end{equation}
Here $\Lcut=\log(1+\Luv^2/m^2)$ and
$\Rcut=\Luv^2/(\Luv^2+m^2)$.
Their zeroth moments give $P_{f\bar f}$ and $S_z^{f\bar f}$, respectively.
Here $S_z^{f\bar f}$ is the constituent spin in a regulated Fock sector.
The real-photon sum rule $\int_0^1 g_1^\gamma(x)\,\dd x=0$
\cite{bass1998photonspin} concerns the physical polarized structure
function, with its factorization prescription and coefficient functions.
Identifying it with the cutoff density $\Delta q^\gamma$ requires
matching these definitions.
The finite GPD terms use the shifted cutoff in Eq.~\eqref{eqS:radial};
the forward orbital moments use $k_\perp^2<\Luv^2$.
The $F_2$ integrals below retain the original relative-momentum disk.
For the logarithmically divergent GPD integrals, changing between disks
in the original and Feynman-shifted momenta changes only terms that
vanish as $\Luv\to\infty$. The bare GPDs retain their ultraviolet
logarithms. The quadratically divergent $F_2$ kernel requires the
original disk and the subtraction specified below.

\section{Derivation of the orbital and spin--orbit moments}
\label{supp:oam}

For $h=+1$ and $k_\perp\ne0$, the overlap expands, up to
$O(\Delta_\perp^2)$, as
\begin{equation}
 W^{[\Gamma]}_{hh}
 =N_c\sum_{s\bar s}w^\Gamma_{s\bar s}A_{s\bar s}^2
 \left[1-i\,\ell_z\frac{\beta(\kb\times\Db)}{k_\perp^2}\right]
 ,
 \label{eqS:expansion}
\end{equation}
where $\Psi^+_{s\bar s}=A_{s\bar s}(x,q_\perp^2)e^{i\ell_z\phi_q}$,
with real $A_{s\bar s}$ after absorbing constant channel phases;
$w=1$ for $\Gamma=\gamma^+$ and $w=s$ for $\Gamma=\gamma^+\gamma_5$.
The azimuths $\phi_\pm$ of $\boldsymbol q_\pm$ obey
$\phi_+-\phi_-=\beta(\kb\times\Db)/k_\perp^2+O(\Delta_\perp^2)$.
At $\Db=\boldsymbol0_\perp$, matching Eq.~\eqref{eqS:helicity} gives
\begin{equation}
 -\frac{k_\perp^2}{m^2}(\FA,\GA)
 =N_c\beta\sum_{s\bar s}(1,s)\,\ell_z A_{s\bar s}^2,
 \label{eqS:pointwise}
\end{equation}
with $\FS(x,\kb,\boldsymbol0_\perp)=N_c\sum A^2$.
Equation~\eqref{eqS:pointwise} identifies the weighted GTMDs at each
$(x,\kb)$ with the wave-function probabilities weighted by the fermion's
orbital share $\beta\ell_z$ and by $s\beta\ell_z$, respectively.
Integrating over $x$ and $\kb$ gives the moments in Letter
Eq.~\Leqref{eq:oamdef}. Pair overlaps use $Z_\gamma=1$ at $O(\alpha)$.

\section{Gravitational moments}
\label{supp:moments}

The fermionic energy-momentum tensor (EMT) moment is
\begin{equation}
 \int_{-1}^1\dd x\,x\,
 H_{h'h}^{[\gamma^+]}(x,0,t)
 =\frac{\langle p',h'|T_f^{++}|p,h\rangle}
 {2(P^+)^2}.
 \label{eqS:localmoment}
\end{equation}
Here $H^{[\gamma^+]}=\int\dd^2k_\perp W^{[\gamma^+]}$ and
$P^+=p^+=p'^+$.
We use Milton's photon tensors~\cite{milton1977photon}, crossed to the
scattering channel and normalized to the positive free-photon
$T^{++}$ matrix element. The tensors multiplying $F_2$ and $F_3$ are
\begin{equation*}
 \begin{aligned}
 \Theta_2^{\mu\nu}&=-(\Delta^\mu\Delta^\nu-tg^{\mu\nu})\\
 &\quad\times[2(\Delta\cdot\epsilon)(\Delta\cdot\epsilon'^*)
 -t(\epsilon\cdot\epsilon'^*)],\\
 \Theta_3^{\mu\nu}&=-4P^\mu P^\nu
 [2(\Delta\cdot\epsilon)(\Delta\cdot\epsilon'^*)\\
 &\hspace{67pt}-t(\epsilon\cdot\epsilon'^*)].
 \end{aligned}
\end{equation*}
Here $P=(p'+p)/2$, $\Delta=p'-p$, $t=\Delta^2$, and
$\epsilon$ and $\epsilon'^*$ are the initial and final photon
polarization vectors.
For $\Db=\Delta_\perp\hat{\boldsymbol x}$, the full EMT obeys
$\langle T^{++}\rangle_{hh}/[2(P^+)^2]=F_1$ and
$\langle T^{++}\rangle_{+-}/[2(P^+)^2]=-2tF_3$.
The equal antifermion moment doubles the fermion term; the one-photon
normalization fixes $F_1(0)=1$ and has no helicity flip.
The resulting finite integrals are
\begin{align}
 \frac{F_1-1}{\af}
 &=2\int_0^1\!\dd x\,x\left[
 (1-c-2c\beta^2\tau)\mathcal S(\beta\sqrt\tau)
 -(1-c)\right],\nonumber\\
 \frac{2tF_3}{\af}
 &=4\int_0^1\!\dd x\,x^2\beta
 \left[1-\mathcal S(\beta\sqrt\tau)\right].
 \label{eqS:finitechecks}
\end{align}
\section{Transverse-EMT overlap for \texorpdfstring{$F_2$}{F2}}
\label{supp:F2}

\subsection{Overlap and forward subtraction}

For on-shell photons, we use the symmetric frame
$\boldsymbol p'_\perp=-\boldsymbol p_\perp=\Db/2$,
with $\Db=Q\hat{\boldsymbol x}$, $t=-Q^2$, and $Q>0$.
The transverse component perpendicular to the transfer gives
\begin{equation}
 \langle p',+|T^{yy}|p,-\rangle=-t^2F_2(t),
 \qquad
 \langle p',+|T^{xx}|p,-\rangle=0.
 \label{eqS:F2projection}
\end{equation}
The minus sign follows from the tensors in Sec.~\ref{supp:moments} and the polarization
vectors used in Letter Eq.~\Leqref{eq:wavefunction}.

The reduced one-fermion overlap kernel is
\begin{equation}
 \begin{aligned}
 \mathcal O_{f;s's}^{yy}(k',k)
 &=\frac{\langle f;k',s'|T_f^{yy}(0)|f;k,s\rangle}
 {\sqrt{k'^+k^+}}\\
 &=\frac{2q_y^2+isq_yQ}{xP^+}\delta_{s's}.
 \end{aligned}
 \label{eqS:F2kerneldef}
\end{equation}
Here $k'^+=k^+=xP^+$, $k'_y=k_y=q_y$, and $s/2$ is the
fermion light-front helicity.
The two orbital channels in Eq.~\eqref{eqS:amplitudes} have equal
helicity-flip wave-function products, so their sum cancels the
$s$-dependent term in the kernel.
The pair overlap, including both the fermion and antifermion, is
\begin{equation}
 \begin{aligned}
 \mathcal M_{\rm pair}^{yy}(Q;\Luv)
 &=2N_cP^+\int_0^1\!\dd x\int_{\Luv}\!\dd^2q_\perp\\
 &\quad\times\sum_{s,\bar s}
 \Psi^{+*}_{s\bar s}(x,\qb')\mathcal O_{f;ss}^{yy}
 \Psi^-_{s\bar s}(x,\qb)\\
 &=\frac{8\af}{\pi}\int_0^1\!\dd y\,y\,K(yQ;\Luv).
 \end{aligned}
 \label{eqS:F2pair}
\end{equation}
Here $\qb'=\qb+(1-x)Q\hat{\boldsymbol x}$, $y=1-x$, and
$\int_{\Luv}\dd^2q_\perp$ denotes the disk
$q_x^2+q_y^2<\Luv^2$ in the original relative momentum.
The antifermion term equals the fermion term after
$x\leftrightarrow1-x$ and $q_x\to-q_x$, giving the factor of two.
Terms odd in $q_y$ integrate to zero on this disk.
The transverse combination is
\begin{equation}
 \begin{aligned}
 K(a;\Luv)&=I(a;\Luv)+aJ(a;\Luv)\\
 &=\int_{\Luv}\!\dd^2q_\perp\,
 \frac{q_y^2(q_x^2-q_y^2+a q_x)}{D_0D_a},
 \end{aligned}
 \label{eqS:F2K}
\end{equation}
where $a=yQ$, $D_0=q_\perp^2+m^2$, and
$D_a=(q_x+a)^2+q_y^2+m^2$.
The two required integrals are
\begin{equation}
 \begin{aligned}
 I(a;\Luv)&=\int_{\Luv}\!\dd^2q_\perp\,
       \frac{q_y^2(q_x^2-q_y^2)}{D_0D_a},\\
 J(a;\Luv)&=\int_{\Luv}\!\dd^2q_\perp\,
       \frac{q_y^2q_x}{D_0D_a}.
 \end{aligned}
 \label{eqS:F2integrals}
\end{equation}
We subtract the forward pair overlap at the same finite cutoff:
\begin{equation}
 F_2(t)
 =-\frac{8\af}{\pi t^2}\lim_{\Luv\to\infty}
 \int_0^1\!\dd y\,y\,[K(yQ;\Luv)-K(0;\Luv)].
 \label{eqS:F2overlap}
\end{equation}
For $m>0$, $F_2(t)$ approaches a finite value as $t\to0$.
Equation~\eqref{eqS:F2projection} therefore requires the transverse
helicity-flip matrix element to be proportional to $t^2=Q^4$ near
zero transfer. The pair overlap contains a transfer-independent forward
contribution, which is canceled in the full EMT.
We implement this cancellation by subtracting $K(0;\Luv)$ at the same
finite cutoff, before dividing by $t^2$ and taking $\Luv\to\infty$.
Here $K(0;\Luv)$ sets the transfer to zero while retaining the integration
over $\qb$.

\subsection{Transverse integrals}

Set $q_x=\sqrt u\cos\phi$, $q_y=\sqrt u\sin\phi$, so that
$\dd^2q_\perp=\dd u\,\dd\phi/2$,
$0\leq u\leq\Luv^2$, and $0\leq\phi\leq2\pi$.
The angular denominator is $D_a=A+b\cos\phi$, where
\begin{equation}
 \begin{aligned}
 A&=u+a^2+m^2,\qquad b=2a\sqrt u,\\
 S&=\sqrt{A^2-b^2}
   =\sqrt{(u+m^2-a^2)^2+4a^2m^2}.
 \end{aligned}
 \label{eqS:F2angularvars}
\end{equation}
The angular identities used are
\begin{equation}
 \begin{aligned}
 \int_0^{2\pi}\!\frac{\dd\phi}{A+b\cos\phi}
 &=\frac{2\pi}{S},\\
 \int_0^{2\pi}\!\frac{\sin^2\phi\,\dd\phi}{A+b\cos\phi}
 &=\frac{2\pi}{A+S},\\
 \int_0^{2\pi}\!\frac{\sin^2\phi\cos\phi\,\dd\phi}{A+b\cos\phi}
 &=-\frac{\pi b}{(A+S)^2},\\
 \int_0^{2\pi}\!\frac{\sin^2\phi\cos^2\phi\,\dd\phi}{A+b\cos\phi}
 &=\frac{\pi A}{(A+S)^2}.
 \end{aligned}
 \label{eqS:F2angles}
\end{equation}
The last three follow by dividing powers of $\cos\phi$ by
$A+b\cos\phi$ and using the first identity.
Since $\sin^2\phi(\cos^2\phi-\sin^2\phi)
=2\sin^2\phi\cos^2\phi-\sin^2\phi$, Eq.~\eqref{eqS:F2integrals}
becomes
\begin{equation}
 \begin{aligned}
 I(a;\Luv)&=-\pi\int_0^{\Luv^2}\!\dd u\,
       \frac{u^2S}{(u+m^2)(A+S)^2},\\
 J(a;\Luv)&=-\pi a\int_0^{\Luv^2}\!\dd u\,
       \frac{u^2}{(u+m^2)(A+S)^2}.
 \end{aligned}
 \label{eqS:F2radial}
\end{equation}
For the radial integrations, use
\begin{equation*}
 \begin{aligned}
 w&=u+m^2-a^2+S,\\
 u&=\frac w2+a^2-m^2-\frac{2a^2m^2}{w},\\
 \dd u&=\left(\frac12+\frac{2a^2m^2}{w^2}\right)\dd w,\\
 S&=\frac w2+\frac{2a^2m^2}{w},
 \qquad A+S=w+2a^2.
 \end{aligned}
\end{equation*}
The integrands are then rational functions of $w$, with limits
$w(0)=2m^2$ and $w(\Luv^2)$.
Integrating gives
\begin{align}
 I(a;\Luv)\simeq\pi\bigg[&
 -\frac{\Luv^2}{4}+\frac{a^2+2m^2}{4}\Lcut
 +\frac{3m^2}{4}+\frac{5a^2}{24}\nonumber\\
 &-\frac{(a^2+2m^2)R_a}{2a}\zeta\bigg],
 \label{eqS:F2I}\\
 J(a;\Luv)\simeq\pi\bigg[&
 -\frac a4\Lcut-\frac a8+\frac {R_a}2\zeta\bigg].
 \label{eqS:F2J}
\end{align}
Here $R_a=\sqrt{a^2+4m^2}$, $\zeta=\operatorname{arsinh}(a/2m)$,
and $\Lcut=\log(1+\Luv^2/m^2)$.
The symbol $\simeq$ retains every divergent and finite term and
omits only terms that vanish as $\Luv\to\infty$ at fixed $a,m$.
Combining the integrals gives
\begin{equation}
 \begin{aligned}
 K(a;\Luv)\simeq\pi\bigg[&
 -\frac{\Luv^2}{4}+\frac{m^2}{2}\Lcut
 +\frac{3m^2}{4}\\
 &+\frac{a^2}{12}
 -\frac{m^2R_a}{a}\zeta\bigg].
 \end{aligned}
 \label{eqS:F2Kasym}
\end{equation}
At zero transfer,
\begin{align}
 K(0;\Luv)
 &=-\frac{\pi}{4}\int_0^{\Luv^2}\!\dd u\,
             \frac{u^2}{(u+m^2)^2}\nonumber\\
 &=-\frac{\pi}{4}
 \left[\Luv^2+m^2\Rcut-2m^2\Lcut\right].
 \rule[-12pt]{0pt}{12pt}
 \label{eqS:F2Iforward}
\end{align}
\noindent Here $\Rcut=\Luv^2/(\Luv^2+m^2)$.
The common terms $-\pi\Luv^2/4$ and $\pi m^2\Lcut/2$
cancel between Eqs.~\eqref{eqS:F2Kasym} and \eqref{eqS:F2Iforward},
giving
\begin{equation}
 \begin{aligned}
 K_{\rm sub}(a)
 &\equiv\lim_{\Luv\to\infty}[K(a;\Luv)-K(0;\Luv)]\\
 &=\pi\left[m^2+\frac{a^2}{12}-\frac{m^2R_a}{a}\zeta\right].
 \end{aligned}
 \label{eqS:F2Jfinite}
\end{equation}
An independent dimensionally regulated evaluation of the conserved
transverse projection gives the same finite kernel:
\begin{equation}
 \begin{aligned}
 K_{\rm sub}(a)
 &=\frac{\pi}{2}\int_0^1\!\dd v\,
 \bigg[v(1-v)a^2\\
 &\qquad
 -m^2\log\!\left(1+\frac{v(1-v)a^2}{m^2}\right)\bigg].
 \end{aligned}
 \label{eqS:F2finiteparameter}
\end{equation}
Expanding Eq.~\eqref{eqS:F2Jfinite} at small $a$ gives
$K_{\rm sub}(a)=\pi a^4/(120m^2)+O(a^6/m^4)$.
Since $a=yQ$, the subtracted overlap is proportional to $Q^4=t^2$
near zero transfer, so dividing by $t^2$ gives a finite $F_2(0)$.
The parallel projection $\langle T^{xx}\rangle_{+-}$ vanishes in
the same conserved-EMT prescription.

Substituting Eq.~\eqref{eqS:F2Jfinite} into
Eq.~\eqref{eqS:F2overlap} gives
\begin{equation}
 \begin{aligned}
 \frac{2tF_2(t)}{\af}
 =\frac4\tau\int_0^1\!\dd y\,\bigg[
 y+\frac{\tau y^3}{3}
 -\frac{\sqrt{1+\tau y^2}}{\sqrt\tau}
       \operatorname{arsinh}(y\sqrt\tau)\bigg].
 \end{aligned}
 \label{eqS:F2integralsum}
\end{equation}
\par\vspace{6pt}
\noindent Here $\tau=Q^2/(4m^2)$, $\eta=\operatorname{arsinh}\sqrt\tau$,
and $r=\sqrt{1+1/\tau}$.
Evaluating the integral gives
\begin{equation}
 \frac{2tF_2(t)}{\af}
 =\frac13+\frac3\tau-\frac{2r\eta}{\tau}
 -\frac{\eta^2}{\tau^2},
 \label{eqS:F2closed}
\end{equation}
with continuous limit $F_2(0)=-\af/(90m^2)$.
\end{document}